# Hardware Complexity Aware Design Strategy for a Fused Logarithmic and Anti-Logarithmic Converter

Botao Xiong and Yuanfeng Sui

*Abstract*—**The logarithmic and anti-logarithmic converters are realized with the piecewise linear approximation method, which is implemented by the shift-and-add architecture. This brief utilizes the similarities of Log and Antilog functions so that the adder tree block and multiplexer block can be shared by the Log and Antilog converters. As a result, the Antilog function can be implemented by the Log converter at the cost of additional 14% area and 6% latency. It implies the shift-and-add architecture can approximate multiple similar nonlinear functions with a slightly hardware cost. In addition, this brief proposes a set of formulas to predict the area and latency of shift-and-add architecture with different quantized coefficients that can facilitate the finding of a trade-off point in the Latency-Area-Precision space.**

## I. Introduction

THE implementation of nonlinear functions can be divided into two categories, analog [1] and digital methods. It is preferable to utilize the digital circuits since the analog circuits suffer from Process-Voltage-Temperature (PVT) variations and the integration of analog circuits and digital circuits increases the system complexity due to Analog-to-Digital Converters (ADC) and Digital-to-Analog Converters (DAC). The digital implementation consists of three solutions: Look-up table (LUT) methods [2], digital recurrence methods [3], and approximation methods that are further divided into two categories: piecewise linear (PWL) approximation [4~8] and piecewise polynomial (PWP) approximation [9~12]. For desired precision level, the PWP reduces the number of regions at the cost of computation efforts. Compared with approximation methods, LUT method [2] and digital recurrence method [3] can reach a high precision level at the cost of area and latency, respectively. For example, a logarithmic (Log) converter based on LUT is proposed in [2] that can achieve 23 bits of fractional precision with 20K bits of ROM.

The Log and Antilog functions can translate several complex operations (e.g. multiplication and division) into several simple arithmetic operations in the logarithmic space [4]. However, in many cases, the Log converter and Antilog converter do not work simultaneously. Hence, if the Antilog converter can share some components of Log converter, the cell area of Antilog converter could be reduced significantly, which motivates this paper to implement a fused Log and Antilog converter. It means the fused converter can perform different functions by choosing different segment index encoders. Since the desired precision level is not high (10-bit), according to the preceding discussion, the PWL is employed in this paper.

The main contribution of this paper is that it is the first time to utilize the Log converter to calculate the Antilog function with a tolerable overhead (14% area and 6% latency). It further proves that the adder tree block and multiplexer block in the shift-and-add architecture can be shared by multiple similar nonlinear functions. Furthermore, this paper proposes a set of formulas to estimate the latency and area of shift-and-add architecture such that a trade-off point is found in the Latency-Area-Precision space. In the end, the proposed design strategy not only controls the maximum error but also takes the error distribution into consideration.

The rest of this paper is organized as follows. Section II introduces a hardware complexity aware algorithm for shift-and-add architecture and discusses the similarity of Log and Antilog functions. Section III presents a design procedure for fused Log and Antilog converter. The hardware implementation is introduced in Section IV.

## II. Basic Idea of Fused Log and Antilog Converter

### A. Shift-and-Add Architecture

The Log and Antilog functions are approximated by the PWL scheme that is implemented by the shift-and-add architecture. To be compatible with the half-precision LNS format, the input and output of fused Log and Antilog converter are 10-bit.

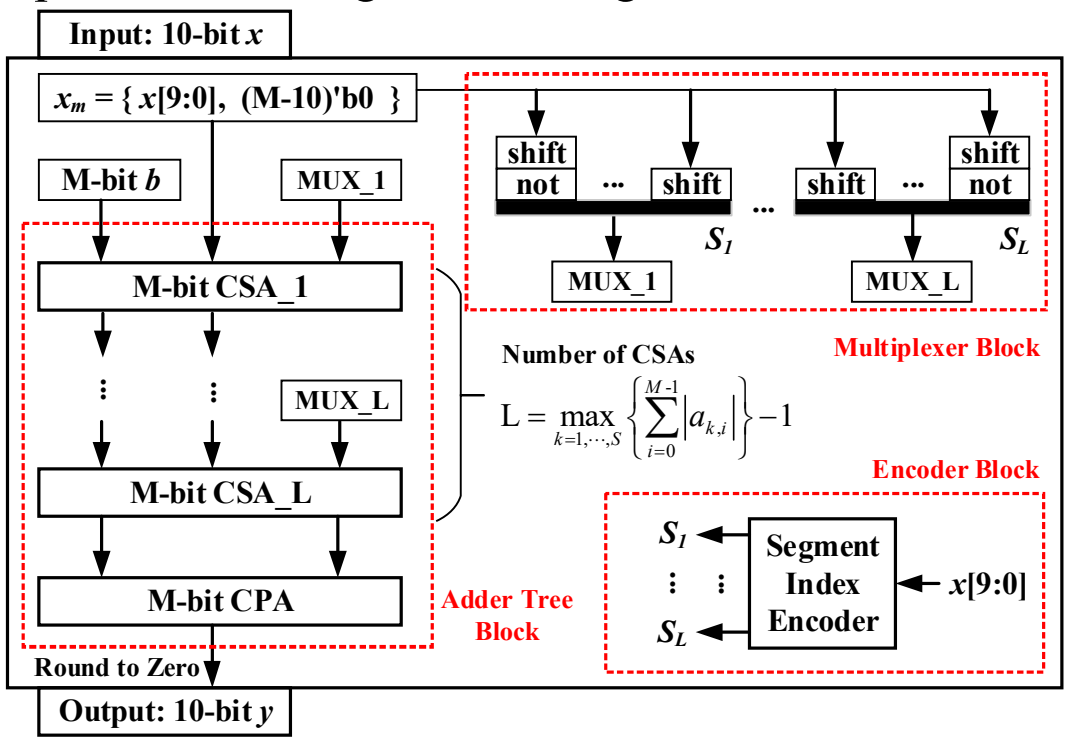

Fig. 1. The basic architecture of piecewise linear approximation methods.

As shown in Fig. 1, a linear function ($ax$+$b$) is implemented by the shift-and-add architecture that consists of an adder tree

Manscript submitted in Nov, 12, 2020. This work was supported in part by the National Natural Science Foundation of China under Grant 61801450 and the Fundamental Research Funds for the Central Universities under Grant DUT20RC(3)058.

Botao Xiong and Yuanfeng Sui are with the School of Microelectronics, Dalian University of Technology, Dalian, 116000, China (e-mail: xiongbotao@dlut.edu.cn, 2412850165@qq.com).

block, a multiplexer block and a segment index encoder block. The adder tree block consists of several carry-save adders (CSA) and a carry-propagation adder (CPA). The advantage of CSA is that the latency of adder tree block is reduced, which is further explained in (1). Since the cost of adder tree block is decided with the nonzero terms of $a$, the canonic signed digit (CSD) scheme is employed. To further reduce the hardware cost, the "add 1" operation of the 2's complement subtraction is ignored. This scheme can remove $M$ one-bit half adders. The reduction ratio is proportion to the number of '-1' of CSD numbers. In addition, the complexity of segment index encoder is proportion to the number of segments that is decided by the segmentation scheme. The complexity of the multiplexer block is related to the encoder block and the $M$-bit CSD numbers $\{a_1 \cdots a_s\}$, where $s$ is the number of segments. This subsection aims to predict the hardware cost of shift-and-add architecture if the coefficients $\{a_1 \cdots a_s\}$ are given.

- **Hardware Cost of Adder Tree Block**

Since the CSA is a set of one-bit full adders without any carry chain, the delay of $M$-bit CSA equals to the delay of one-bit full adder. Similarly, the CPA is a set of one-bit full adders with a carry chain. The delay of $M$-bit CPA is $M$ times longer than the delay of one-bit full adder. Therefore, the latency of the adder tree block is written as

$$\left\{ \max_{k=1,\cdots,S} \left\{ \sum_{i=0}^{M-1} \left| a_{k,i} \right| \right\} - 1 \right\} + M \tag{1}$$

where $a_{k,i}$ is the *i-th* bit of $M$-bit CSD number $a_k$ for the *k-th* segment. The first term of (1) is the number of CSAs. In addition, the area of the adder tree block is represented as the number of one bit full adders written as

$$\max_{k=1,\cdots,S} \left\{ \sum_{i=0}^{M-1} \left| a_{k,i} \right| \right\} \times M \tag{2}$$

- **Hardware Cost of Multiplexer Block**

The complexity of multiplexer block is approximated by the number of right shift operators. In this design, the $M$-bit CSD number $a$ includes one integer bit and ($M$-1) fractional bits. In addition, the integer bit is always equal to one that does not require any shift operations. Therefore, the number of right shift operators is written as

$$\sum_{i=1}^{M-1} \left[ \max_{k=1\cdots S} \left\{ a_{k,i} \right\} - \min_{k=1\cdots S} \left\{ a_{k,i} \right\} \right] \tag{3}$$

- **Hardware Cost of Encoder Block**

The outputs of encoder are the select lines of multiplexer. In addition, the outputs of encoder of Log converter differ from the outputs of encoder of Antilog converter. Hence, the encoder block cannot be shared. However, this block makes up only a small percentage of area of Log or Antilog converter.

### *B. Similarity of Log and Antilog Functions*

The $M$-bit CSD number $a$ is related to the slope of a curve. As shown in Fig. 2, the first derivatives of Log and Antilog functions are almost symmetric with respect to a line ($x$=0.5). The red line ( Flip the first derivative of Antilog function about the line ($x$=0.5) and then calculate the difference ) implies the CSD numbers of the first several segments of Log converter are almost equal to the CSD numbers of the last several segments of Antilog converter. It implies the adder tree and multiplexer block could be shared by the Log and Antilog converter.

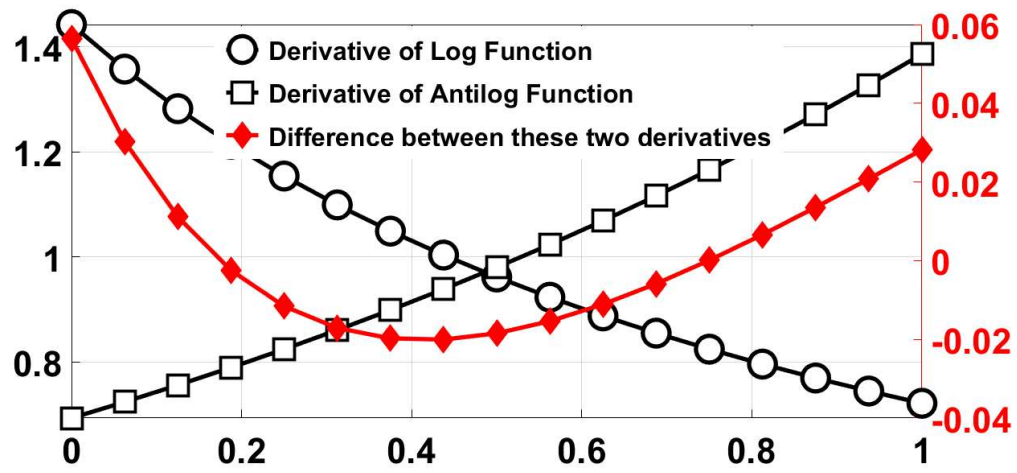

Fig. 2. The similarity of Log ($\log_2(1+x)$) and Antilog ($2^x$-1)functions.

## III. Design Flow of Fused Log and Antilog Converter

In general, the PWL method includes four steps. (1) Define an objective function. (2) Divide the interval into several regions. (3) Select a linear function to estimate the nonlinear function in each region. (4) Quantize the coefficients. In order to integrate the Log converter and Antilog converter together, the step 4 is further divided into two sub-steps. First, construct a space including all feasible CSD numbers. Then, select a set of optimum CSD numbers such that the adder tree block and multiplexer block can be reused, which are discussed in Section III.C and Section III.D, respectively.

### *A. Step 1 : Objective Function*

A nonlinear function $f(x)$ is approximated by a polynomial of degree $n$ denoted as $P_n(x)$. Then, the maximum hardware error (MHE) is defined as

$$\mathbf{MHE} = \max_i \left| \mathbf{HW}\left[ P_n(x_i) \right] - \mathbf{RN}\left[ f(x_i) \right] \right| \tag{4}$$

where $\mathbf{HW}(\cdot)$ is referred to as the hardware implementation of polynomial function, $\mathbf{RN}(\cdot)$ is the round to nearest operator and the subscript $i$ is the index of the $N$-bit binary number $x_i$. As a result, the approximation error can be written as (5) that is consistent with the round-off error. The approximation error propagation is analyzed with the interval analysis method.

$$f(x_i) \in \mathbf{HW}\left[ P_n(x_i) \right] \pm (\mathbf{MHE} + 0.5) \times \mathbf{LSB} \tag{5}$$

The desired MHE in this paper equals to one. Therefore, the objective function is defined as

$$\mathbf{Minimize} \quad N_{+1} + N_{-1} + \left| N_{+1} - N_{-1} \right| \tag{6}$$

where $N_{+1}$ is the number of data points whose hardware error defined as (4) is equal to +1; $N_{-1}$ is the number of data points whose hardware error is equal to -1.

Compared to the maximum error range [4], maximum relative error [5], error equal distribution [6], and signal-to-noise ratio [7], this objective function (6) not only controls the maximum absolute error (MAE<1.5×LSB) but also enables the absolute error has a symmetric distribution and zero expectation, which is proved in Section IV (Tab III, Tab IV and Fig.5).

It is difficult to calculate the MHE since $\mathbf{HW}(\cdot)$ is unknown. However, the bound of MHE is calculated by the maximum quantization distance (MQD) written as

$$\mathbf{MQD} = \max_i \left| P_n(x_i) - \mathbf{RN}\left[ f(x_i) \right] \right| \tag{7}$$

If MQD < 0.5×LSB, the polynomial is able to reach the level of accuracy of ideal Log or Antilog converter. The MHE could be reduced to zero with proper hardware design. Similarly, the MHE can be reduced to one if 0.5×LSB<MQD<1.5×LSB. In general, the bound of MHE is equal to **RN**[MQD/LSB]. Hence, the coefficients of approximation polynomial are solved by

$$\begin{aligned} &\min_{a_k^* \in R} \quad \mathbf{MHE} = \mathbf{RN}\left[\mathbf{MQD/LSB}\right] \\ &s.t. \quad \left| \sum_{k=0}^{n} a_k^* x_i^k - \mathbf{RN}\left[ f\left(x_i\right)\right] \right| < \mathbf{MQD} \end{aligned} \tag{8}$$

*B. Step 2: Segmentation Scheme*

The interval [0, 1] should be divided into several nonuniform or uniform regions such that the desired MQD can be fulfilled with a proper approximation polynomial in each region.

- **Recursive Segmentation Scheme**

First, the interval [0, 1] is divided into two uniform segments. The MQDs of each segment can be solved with (8). Then, the segment, whose MQD is larger than 1.5×LSB, should be further divided into two sub-segments. Repeat this process until all constraints are satisfied. The advantage of this method is that the encoder is simply implemented by several multiplexers.

- **MQD-Flattened Segmentation Scheme**

The drawback of recursive segmentation scheme is that the distribution of MQDs on each region is not uniform. Hence, a MQD-Flattened segmentation is proposed. Because the MQD is related to the length of the segment, reduce the length of the segment ($k_{max}$) that has a maximum MQD and increase the length of the segment ($k_{min}$) that has a minimum MQD. Repeat this process to minimize the range of MQD.

```
Algorithm: MQD-Flattened Segmentation Scheme
Input:    { (x_i, y_i) | i = 0,···,1023}
Output:   Segment_Index_Encoder
      // The interval [0, 1] is divided into 64 uniform sub-segments.
      // Each sub-segment consists of 16 data points.
      // Region(k): Number of sub-segments in the k-th region
      Initialization: Region(k),  k=1,2, ···,S
      // Utilize (8) to calculate the MQD of each region
      for k=1:S
          MQD(k) = LinearProg(Region(k));
      endfor
      // Adjust the length of Region(k_max) and Region(k_min)
      while( max(MQD) – min(MQD) > Threshold )
         k_max = find ( MQD == max(MQD) );
         k_min  = find ( MQD == min (MQD) );
         Region(k_max) = Region(k_max) – 1;
         Region(k_min ) = Region(k_min ) + 1;
         for k=1:S
              MQD(k) = LinearProg(Region(k));
         endfor
      endwhile
```

*C. Step 3: Generate the Latency-Area-Precision Space*

If the $M$-bit CSD numbers $\{a_1 \cdots a_S\}$ are given, the prediction of hardware complexity of shift-and-add architecture has been discussed in section II. Moreover, this subsection constructs a space including all feasible CSD numbers $\{a_1 \cdots a_S\}$ and binary numbers $\{b_1 \cdots b_S\}$ satisfying the constraint (MHE=1).

As shown in Fig. 3, specify the desired MQD and then utilize the proposed segmentation schemes to divide the interval [0, 1] into several segments. Meanwhile, the optimum coefficients of each segment are obtained.

The optimum coefficients $a^*$ and $b^*$ are further quantized to the CSD numbers and binary numbers, respectively. First, the quantization bit number $m$ is specified. For the $k$-th segment, a $m$-bit CSD number $a[k,j,m]$ is sampled from the neighborhood of the optimum coefficient $a_k^*$. Second, the $m$-bit binary number $b[k,j,m]$ is calculated by minimizing error (6). In the end, the latency and number of adders are calculated by (1) and (2).

In summary, the Latency-Area-Precision space is expressed as a set of matrixes with different quantization bit numbers $m$.

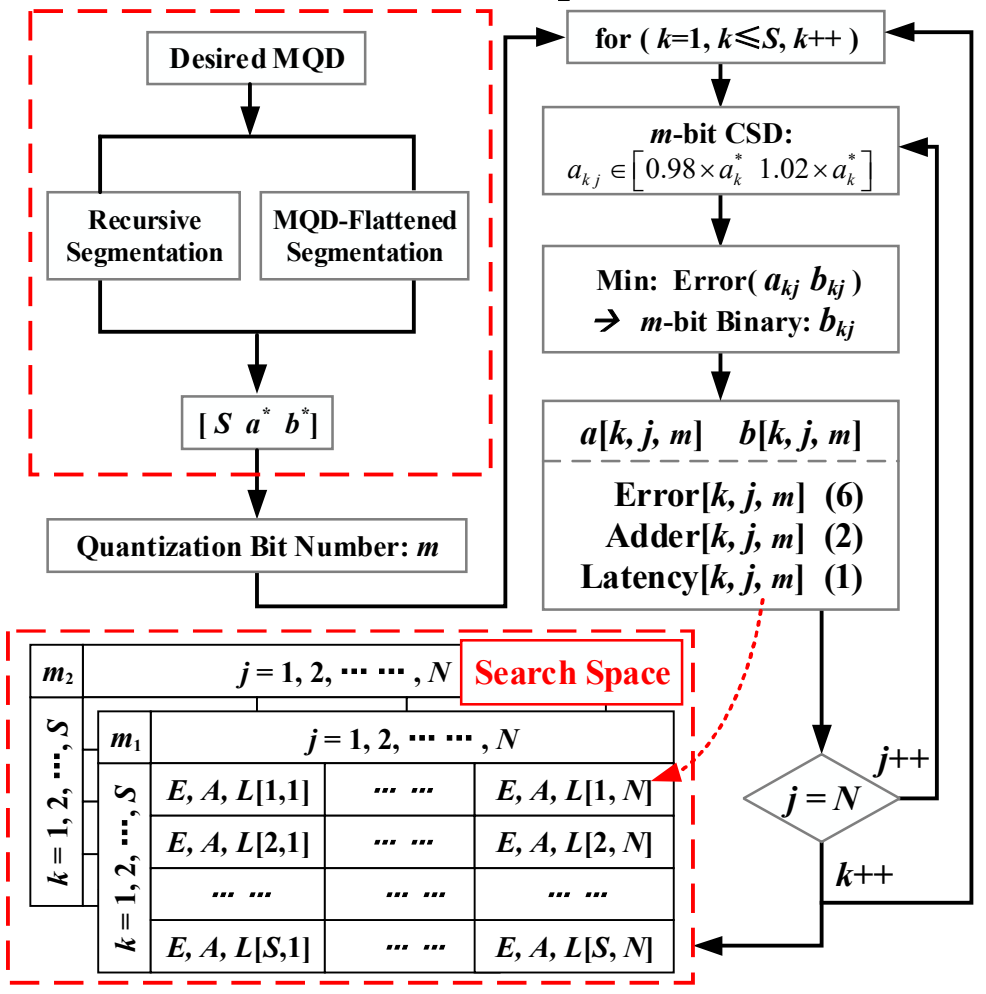

Fig. 3. Construct the Latency-Area-Precision space.

*D. Step 4: Search in the Latency-Area-Precision Space*

This subsection discusses how to find a trade-off in the search space. If the quantization bit number $m$ is given, the minimum error is simply estimated by

$$\sum_{k=1}^{S} \min_{1\le j\le N} \left\{ \mathrm{Error}\left[k,j,m\right] \right\} \tag{9}$$

In addition, the minimum latency and minimum number of one-bit full adders of adder tree block are written as

$$\begin{cases} \max\limits_{1\le k\le S} \left\{ \min\limits_{1\le j\le N} \mathrm{Latency}\left[k,j,m\right] \right\} \\ \max\limits_{1\le k\le S} \left\{ \min\limits_{1\le j\le N} \mathrm{Adder}\left[k,j,m\right] \right\} \end{cases} \tag{10}$$

According to (1) and (2), the minimum latency and number of adders are proportion to the minimum number of CSAs that is written as

$$\max_{1\le k\le S} \left\{ \min_{1\le j\le N} \left\{ \sum_{i=0}^{m-1} \left| a\left[k,j,m\right]\left[i\right] \right| -1 \right\} \right\} \tag{11}$$

The impacts of quantization bit number $m$ on the minimum number of CSAs and minimum error are shown in Fig. 4, which reveals that the increasing $m$ not only cannot further improve the precision, but also requires more hardware resources. Hence, the $m$ is set to be 14 or 13 for the recursive segmentation scheme or MQD-Flattened segmentation scheme, respectively.

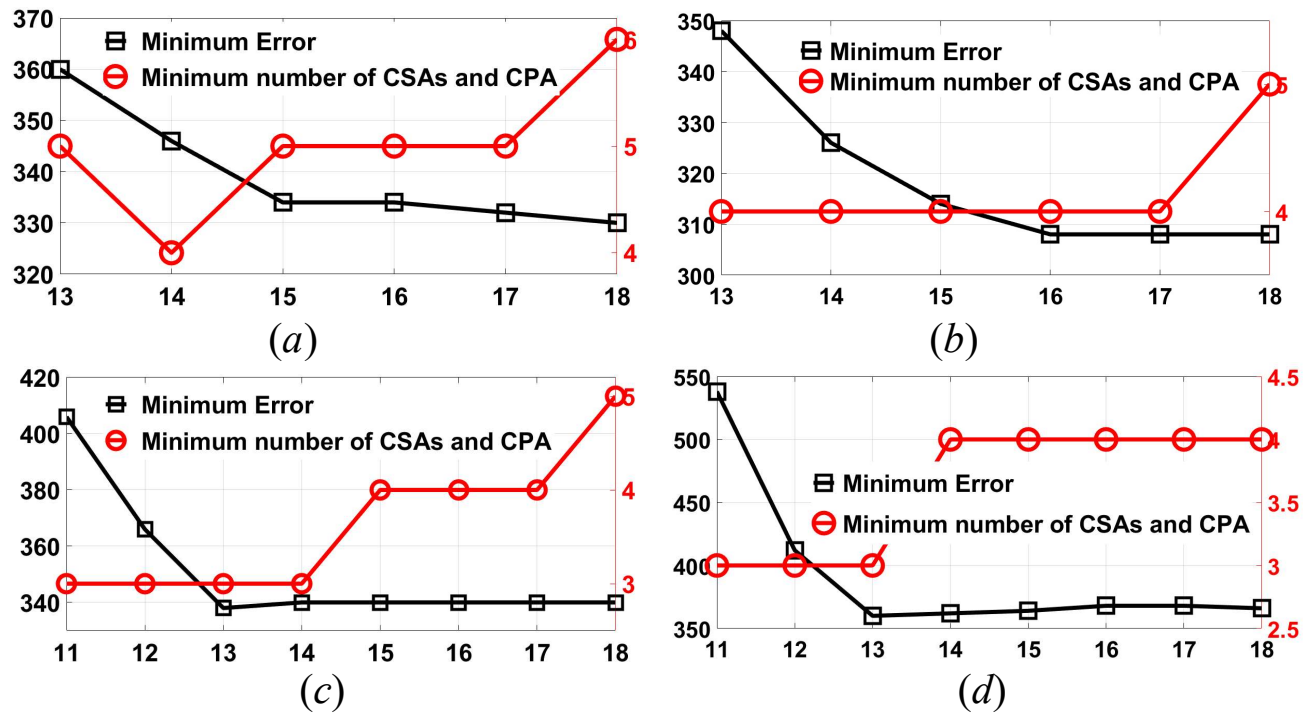

Fig. 4. The impact of quantization bit number $m$ on the minimum number of CSAs and minimum error. ($a$) and ($b$) are the Log and Antilog converters with recursive segmentation. ($c$) and ($d$) are the Log and Antilog converters with MQD-Flattened segmentation. The $x$-axis is the quantization bit number $m$.

Once the quantization level $m$ is determined, the search space is reduced from several matrixes to a single matrix. Two search schemes are proposed, one aims to reach the minimum error, the other one aims to reach the minimum hardware resources.

- **Search Algorithm: Minimum Error**.

**S1.** For the $k$-th segment, remove the elements whose error are larger than $\min\{\text{Error}(k, :, M)\}$. Repeat this process such that the minimum error defined as (9) is converged.

**S2.** Utilize (10) to calculate the minimum number of adders of the elements that have not been deleted. Then, remove all elements requiring more adders.

**S3.** Select an element in each row from the rest of the elements. Then, a feasible solution $\{a[1, j_1, m], \cdots, a[s, j_s, m]\}$ is obtained and the number of shift operators is calculated by (3). Then, choose a feasible solution with minimum number of right shift operators.

- **Search Algorithm: Minimum Hardware Resources**.

**S1.** Utilize (10) to calculate the minimum number of adders. Then, remove all elements requiring more adders.

**S2.** Select an element in each row from the rest of the elements such that the number of right shift operators is reduced that yields a feasible solution $\{a[1, j_1, m], \cdots, a[s, j_s, m]\}$.

**S3.** Calculate the error defined in (6) of this solution. If the precision is lower than the requirement, go back to the step 2.

### *E. Results of Proposed Design Strategy*

The search algorithm for minimum resources is used to select a trade-off in the Latency-Area-Precision space. The results are shown in Tab. I and Tab. II, where the CSD number $a$ marked by (1, -3) equals to $1+2^{-1}-2^{-3}$.

TABLE I
COEFFICIENTS OF FUSED CONVERTER FOR RECURSIVE SEGMENTATION

| Logarithmic Converter | | | Anti-logarithmic Converter | | |
|---|---|---|---|---|---|
| **10-bit *x*** | ***a*** | ***14-bit b*** | **10-bit *x*** | ***a*** | ***14-bit b*** |
| 0 ~ 63 | 1, -3, 6 | 20 | 0 ~ 127 | -2, -5, 8 | 0 |
| 64 ~ 127 | 2, 4, 6 | 83 | 128 ~ 255 | -2, 5, 8 | 16260 |
| 128 ~ 255 | 2, -5, -8 | 323 | 256 ~ 383 | -3, -6 | 15962 |
| 256 ~ 383 | 3, -5, 8 | 808 | 384 ~ 511 | -4, 8 | 15456 |
| 384 ~ 511 | 6, -8 | 1328 | 512 ~ 639 | 6 | 14852 |
| 512 ~ 639 | -4, -6 | 2056 | 640 ~ 767 | 3, -6, 8 | 13863 |
| 640 ~ 767 | -3, -6 | 2692 | 768 ~ 831 | 2, -5, -8 | 12619 |
| 768 ~ 895 | -2, 4, -6 | 3454 | 832 ~ 895 | 2, -8 | 12189 |
| 896~1023 | -2, -8 | 4179 | 896~1023 | 2, 4, 6 | 10998 |

TABLE II
COEFFICIENTS OF FUSED CONVERTER FOR MQD-FLATTENED SEGMENTATION

| Logarithmic Converter | | | Anti-logarithmic Converter | | |
|---|---|---|---|---|---|
| **10-bit *x*** | ***a*** | ***13-bit b*** | **10-bit *x*** | ***a*** | ***13-bit b*** |
| 0 ~ 95 | 1, -3 | 13 | 0 ~ 159 | -2, -6 | 8188 |
| 96 ~ 191 | 2, 6 | 98 | 160 ~ 287 | -2, 4 | 8085 |
| 192 ~ 303 | 3, 5 | 269 | 288 ~ 431 | -3, 6 | 7898 |
| 304 ~ 415 | 4 | 501 | 432 ~ 575 | -5 | 7625 |
| 416 ~ 559 | -6 | 763 | 576 ~ 703 | 4, 6 | 7123 |
| 560 ~ 703 | -3, 6 | 1178 | 704 ~ 815 | 3, 5 | 6680 |
| 704 ~ 847 | -2, 4 | 1622 | 816 ~ 943 | 2, 6 | 5964 |
| 848~1023 | -2 | 2055 | 944~1023 | 1, -3 | 5131 |

## IV. HARDWARE IMPLEMENTATION

The absolute error of different methods are shown in Tab. III, which proves the average error of Tab. I and Tab. II are better than counterpart works.

TABLE III
STATISTICAL PROPERTIES OF ABSOLUTE ERROR OF DIFFERENT METHODS

| Method | Output | Regions | Mean (error) | Std (error) |
|---|---|---|---|---|
| Ref [6] Log | 26-bit | 11 | $-4.230\times10^{-4}$ | $6.471\times10^{-4}$ |
| Ref [7] Log | 12-bit | 7 | $-2.163\times10^{-4}$ | $5.657\times10^{-4}$ |
| Ref [8] Log | 27-bit | 9 | $-9.776\times10^{-4}$ | 0.0011 |
| Ref [8] AntiLog | 27-bit | 8 | $4.817\times10^{-4}$ | $5.015\times10^{-4}$ |
| Tab.I. Log | 10-bit | 9 | $-2.722\times10^{-5}$ | $4.716\times10^{-4}$ |
| Tab.II.Log | 10-bit | 8 | $-1.035\times10^{-4}$ | $5.372\times10^{-4}$ |
| Tab.I. AntiLog | 10-bit | 9 | $3.383\times10^{-6}$ | $4.847\times10^{-4}$ |
| Tab.II.AntiLog | 10-bit | 8 | $-1.207\times10^{-4}$ | $5.902\times10^{-4}$ |

In addition, the precision of counterpart works are measured with many criteria (e.g. maximum positive/ negative error or relative error or error range ). Because of the page limitation, the absolute error distributions of different works are shown in Fig. 5 that can provide more detailed information. For example, the absolute error distributions of Tab. I and Tab. II are almost symmetric with respect to a line (error = 0) that implies that the average error is very close to zero. In addition, the maximum positive/negative absolute error of different methods also can be found in Fig. 5.

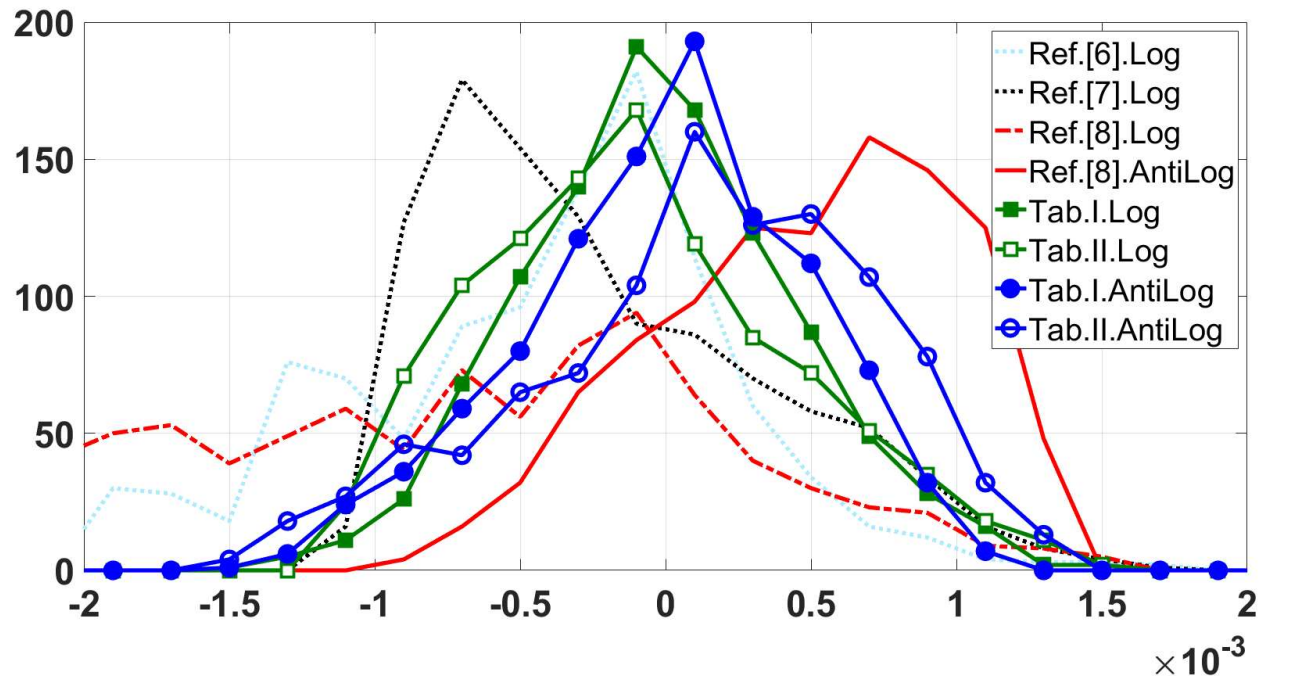

Fig. 5. The error distribution of different methods. The $y$-axis is the number of data points. The $x$-axis is the absolute error.

If the Log or Antilog converter is used in the half-precision LNS, the outputs of [6~8] should be truncated to 10-bit by Round-to-Zero (RZ) or Round-to-Nearest (RN) mode. If the RZ mode is chosen, the hardware architecture of [6~8] does not change. If the RN mode is chosen, a round unit (10 half adders) is required that slightly increases the hardware cost. As shown in Tab. IV, the MHE of [6~8] are larger than one. The proposed methods have a symmetric error distribution.

TABLE IV

THE MAXIMUM HARDWARE ERROR OF DIFFERENT METHODS

| Error (4) | < -2 | -2 | -1 | 0 | 1 | 2 | >2 |
|---|---|---|---|---|---|---|---|
| [6] Log RZ | 0 | 212 | 526 | 266 | 19 | 1 | 0 |
| [6] Log RN | 0 | 114 | 319 | 508 | 78 | 5 | 0 |
| [7] Log RZ | 0 | 49 | 568 | 346 | 61 | 0 | 0 |
| [7] Log RN | 0 | 0 | 296 | 545 | 171 | 12 | 0 |
| [8] Log RZ | 234 | 225 | 325 | 206 | 34 | 0 | 0 |
| [8] Log RN | 151 | 233 | 243 | 319 | 73 | 5 | 0 |
| [8] AntiLog RZ | 0 | 4 | 210 | 582 | 228 | 0 | 0 |
| [8] AntiLog RN | 0 | 0 | 448 | 440 | 498 | 38 | 0 |
| Tab.I. Log | 0 | 0 | 170 | 709 | 145 | 0 | 0 |
| Tab.II.Log | 0 | 0 | 261 | 607 | 156 | 0 | 0 |
| Tab.I. AntiLog | 0 | 0 | 164 | 692 | 168 | 0 | 0 |
| Tab.II.AntiLog | 0 | 0 | 171 | 556 | 297 | 0 | 0 |

The architecture of fused Log and Antilog converter (Tab. II) is shown in Fig. 6. The Log or Antilog converters are switched by choosing different encoders that can be simply implemented by a multiplexer. The novelties of proposed architecture are: (1) the adder tree block is shared by the Log and Antilog converter; (2) add a right shift operator followed by an inverter ($\overline{>>}$ 5, blue color) such that the multiplexer block of the Log converter can be reused by the Antilog converter.

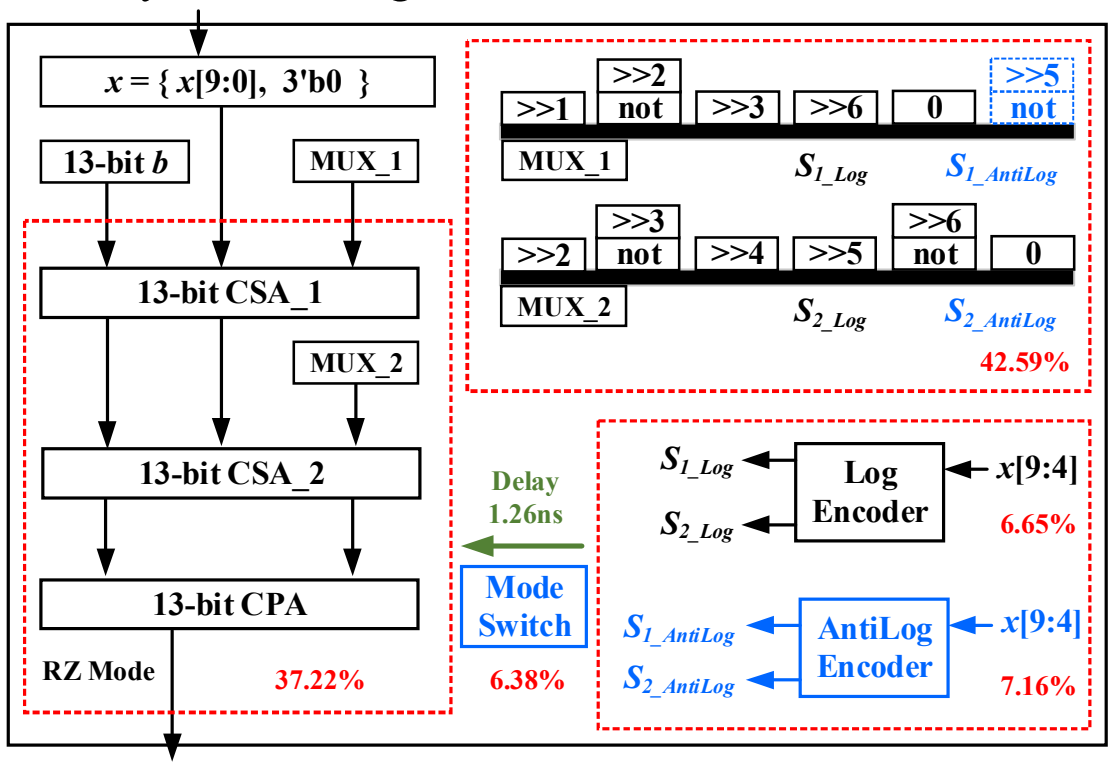

Fig. 6. The architecture of fused converter (Tab. II).

Compared with the Log converter (black color), the Antilog function is calculated by adding two blocks: 'Antilog Encoder' and 'Mode Switch' that occupy about 7.2% and 6.4% cell area. Moreover, the Antilog Encoder and Log Encoder are parallel structure. The latency of these two encoders are almost equal. Therefore, the impact of the Antilog Encoder on the latency of fused converter is negligible. On the other hand, the Antilog and Log encoder are never active simultaneously that is beneficial for low power design. Additionally, the 'Mode Switch' block clearly contributes the rise of the latency of fused converter about 6%. These data come from the Design Compiler based on SMIC.18 Technology.

The hardware cost of different methods are shown in Tab. V. In accordance with Section II.A, the complexity of encoder is related with the number of regions shown in the first line. The complexity of adder tree is depicted by the CSA and CPA. The complexity of multiplexer block is estimated by the number of shifters. In addition, the area, latency, and power (working at 10MHz) of different methods are shown in the following lines. As shown in [8], the cell area of Log converter is almost equal to the Antilog converter. However, by sharing the adder tree block and multiplexer block, the Antilog converter can be easily realized with additional 14% cell area.

TABLE V

SYNTHESIS RESULTS BASED ON SMIC.18 TECHNOLOGY

| | [6] Log | [7] Log | [8] Log | [8] Anti | Tab.I | | Tab.II |
|---|---|---|---|---|---|---|---|
| Regions | 11 | 7 | 9 | 8 | 9 | | 8 |
| CSA | 3×26 | 2×12 | 3×27 | 2×27 | 3 × 14 | | 2 × 13 |
| CPA | 1×26 | 1×12 | 1×27 | 1×27 | 1 × 14 | | 1 × 13 |
| Shifter | 13 | 10 | 8 | 10 | 13 | | 10 |
| Area ($um^2$) | 1335 | 587 | 1235 | 1025 | Log | 917 | 667 |
| | | | | | Fused | 1040 | 767 |
| Latency (ns) | 33.85 | 18.73 | 34.99 | 34.67 | Log | 19.1 | 18.9 |
| | | | | | Fused | 20.2 | 20.16 |
| Power 10MHz | 128.8 uW | 60.2 uW | 111.6 uW | 103.6 uW | 92.1 uW | | 74.9 uW |

## V. CONCLUSION

By using the similarities of several nonlinear functions, the adder tree and multiplexer blocks of shift-and-add architecture can be shared such that the hardware complexity is significantly reduced. The hardware complexity aware technology is able to find a trade-off in the Latency-Area-Precision space.